# PATTERN LANGUAGES AS MEDIA FOR THE CREATIVE SOCIETY

Takashi Iba

Faculty of Policy Management, Keio University
Endo 5322, Fujisawa
Kanagawa, Zip 252-0882, Japan
e-mail: iba@sfc.keio.ac.jp

## ABSTRACT

This paper proposes new languages for basic skills in the Creative Society, where people create their own goods, tools, concepts, knowledge, and mechanisms with their own hands: the skills of learning, presentation, and collaboration. These languages are written as a pattern language, which is a way of describing the tacit practical knowledge. In this paper, a new type of pattern languages are proposed as "Pattern Languages 3.0" and three examples are introduced: Learning Patterns, Collaboration Patterns, and Presentation Patterns. By analyzing the functions with the social systems theory and the creative systems theory, pattern languages are considered as communication media and discovery media.

## INTRODUCTION

When trying to imagine what kind of a society will come to be, we often look at the transition between the past, present, and the future. We look at the flow between the past and the present, and then extend the flow to gain an image of the future.

One way to look at the past and the present of our current world is the transition from a society of "Consumption" to a society of "Communication." Many of what used to be just end consumers receiving the product have now become senders of their own information.

Taking this context into consideration, I see the next generation to be full of people each creating things for their own good. After the transition from a time of mass receiving to personal sending, comes the transition from just plain communication of already information to a time of creation. In this paper this new society to come will be called the Creative Society [1](Figure 1).

---

[1] The term *Creative Society* was coined by Mitchel Resnick in his paper (Resnick, 2007). Also, there is similar term *Creative Economy* by Richard Florida (Florida, 2002).

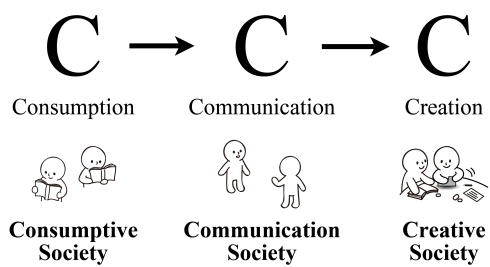

*Figure 1: Transition of our society*

This paper will first introduce how the society will change with the coming of the Creative Society. In its following sections, I will describe how Pattern Languages are useful as media that support people become creative in such times.

## THE COMING OF THE CREATIVE SOCIETY

People of a Creative Society create their own goods, tools, concepts, knowledge, mechanisms, and ultimately the future with their own hands. Creation in this society is no longer limited to just companies and organizations, but is entrusted to the individuals.

P.F. Drucker insisted on the arrival of a somewhat similar idea of a Knowledge Society, but his theory was centered around the social changes that would come in the human-company relations (Drucker, 1993). There is no doubt that social innovation will become a key factor as in the Knowledge Society, but the Creative Society is more radical in the aspect that it handles creation as an event not limited within the enterprises, separate from work labors.

Therefore, the coming of the Creative Society must be viewed not just as a change in how companies and laborers operate, but as a dynamism of the society as a whole. Open-source software developers and Wikipedia page editors have formed communities in a way not seen in usual human communities. What can be seen from this is the fact that developments in informational technologies are now allowing humans to connect and form communities with other people around the world freely.

Similarly, a Creative Society opens up the act of creation previously closed to companies and organizations to the general public. People will be networking with other people through the open act of creation, and everywhere creative communities will be forming. Therefore, what used to be just companies doing the creation will shift towards the idea of communities forming through the process of collaborative creation.

What must not be misunderstood, is that the coming of the Creative Society does not mean that companies and organizations will no longer have its significance. However, it is also true that they must start searching for new ways to cooperate with the emerging creative communities formed by the people. Conventional relationships between individuals and companies were limited to individuals giving labor to the companies, or the companies supplying goods or services to the people. This structure will most likely change in the Creative Society, where collaborative partnerships will be formed between the companies and the individuals.

This new relation in the society will consequently change how companies and organizations must handle the information and knowledge that accumulates within. With people both inside and outside the company freely collaborating, it has simply become impossible to seize control of information. Companies must notice that confining information within themselves will only disadvantage them, and progress of the creative process will be pulled back.

With these problems abiding, we are faced with the need for a way to scribe out and share the knowledge accumulated in the company. Transferring knowledge concerning creativity will especially become increasingly important.

As the consumptive society had media to support its process of production, distribution, and sales, and the communicative society had information media that supported the transfer of information, the Creative Society too must have media that meets its needs. This new kind of media that supports the creative acts of the people living in a Creative Society will be called discovery media.

Discovery media will exist to trigger the creativity of people, and then amplify it. It must be effective not only for individuals, but also for groups of people, and ultimately empower the whole society.

Pattern Languages are what I see as the perfect fit for such a discovery media. A Pattern Language is a language consisted of Patterns, which together scribe out the practical knowledge related to creativity.

## PATTERN LANGUAGES OF CREATIVE ACT

The original idea of pattern languages for writing out the design knowledge was proposed by an architect Christopher Alexander. His intention was to have people get involved in the process of designing their own houses and community where they live in. Ten years later, the idea of pattern languages was adopted in the field of software design. And, recently, the fields where pattern languages are applied are expanding little by little.

With these events in the background, I have been writing pattern languages in a whole new area of knowledge: human actions, like learning, presentation, collaboration, education, business, social innovation, policymaking, and even beauty in daily life. Through these experiences, I have faced such a fundamental question as "what are pattern languages?" Such thoughts allowed me the opportunity to organize and refine my views on pattern languages.

Human actions appear to be much different from architecture and software, however they need to be designed with tacit design knowledge, which consists mainly of context, problem, and solution. In that sense, without losing the essence of design knowledge, pattern languages have gone through, and are still experiencing, a tremendous evolution.

In what follows, I show a new wave around pattern languages, a method to describe the design knowledge in a certain domain of profession. (Alexander 1964, 1979; Alexander et. al. 1977). We will call the new stage the Pattern Language 3.0 stage, distinguishing it from the previous stages (Iba 2011a): Pattern Language 1.0 and Pattern Language 2.0. In what follows, I will present the evolution of pattern languages and describe the difference among these stages (Figure 2). In order to understand the characteristics of each of the stages, the following three aspects will be considered: object of design, act of design, and purpose (Figure 3).

### Three Generations of Pattern Languages

Pattern languages are used to scribe out the design knowledge that lurks in an area of profession. Design knowledge refers to both the intelligence of noticing problems, and solving it. Pattern languages in other words, describe the expertise on the problems that occur under certain contexts, and it's solutions or actions that can be taken to solve the problem.

The original idea of using pattern languages to write out the design knowledge was proposed by an architect Christopher Alexander. His intention was to have the people get involved in the process of designing their own cities and homes that they live in. The late 1970's book, which he wrote with his colleagues, contained 253 patterns on practical architectural design.

Ten years passed since the book was published, and Alexander's idea of pattern languages was adopted in the field of software design. By the 1990's, pattern languages were widely known and used in the field.

With these events in the background, I have been writing pattern languages in a whole new area of knowledge: human actions. Through the experience, I have faced such fundamental questions as "what are pattern languages?" or more practical questions like "how can pattern languages help people?" Such thoughts allowed me for an opportunity to organize and refine my views on pattern languages. This paper will present a new framework for pattern languages based on these thoughts, for future discussions in the field.

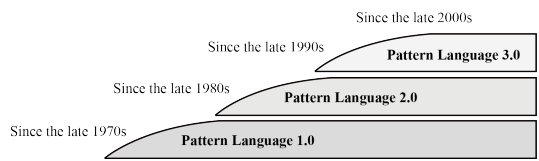

*Figure 2: Three Generations of Pattern Languages*

### A New Object of Design

Pattern language, as stated above, is a language used to write out the design knowledge that lies in an area of profession. The areas of profession that the pattern languages were used for has changed over time. In the first Pattern Language 1.0, its primary object of design was architecture. In contrast to Patten Language 1.0 dealing with material objects, Pattern Language 2.0 focuses on program codes of software. This shift to a nonmaterial subject is a considerable difference that distinguishes the two stages. The evolution then goes on to dealing with human actions, such as innovation, education, learning, collaboration, and presentation, in the new Pattern Language 3.0.

As an example of a pattern language dealing with human actions from the 3.0 stage, I would like to give the Learning Patterns that I have made with my students. The patterns describe the tips, or the design knowledge, for a creative and autonomous style of learning. Patterns from the language are written in short and simple sentences with a picture to describe it. These booklets of the Learning Patterns have been handed out and used by every student at the Shonan-Fujisawa Campus (SFC) of the Keio University since 2009. The patterns are also available for anyone who is learning through its distribution via the web and Twitter. The Learning Patterns is a perfect example of using pattern languages to help design human actions. Other pattern languages exist for presentation (Iba, et. al., 2012), education (Anthony, 1996; Bergin, 2000; Koppe, 2011; Koppe and Nijsten, 2012a; Koppe and Nijsten, 2012b; Larson, Trees, and Weaver, 2008; Bergin, Eckstein, Manns, and Wallingford, 2011; Bergin, Eckstein, Manns, and Sharp, 2011; Bergin, Eckstein, Manns, and Sharp, 2011), change making (Shimomukai, et.al., 2012), life design (Arao, et.al., 2012), and policy making, all of which are a part of the 3.0 stage.

The most important feature of the Pattern Language 3.0 stage must be the fact that the one who designs is also simultaneously the one being designed. In other words, the users of the 3.0 patterns would be using them to design the actions of themselves. This self-referential structure is a characteristic not found in the previous stages of pattern languages. The metacognition resulting from users constantly reflecting on themselves plays a big role in the use of this new kind of patterns.

### A New Act of Design

I've stated above that pattern languages are written about the design knowledge of a field of profession. Though its object of design has changed over time, the meaning of design itself has too changed over the three stages.

The architectural process in the Pattern Language 1.0 stage can be divided into two phases: the stage where buildings and towns are planned out and actually built, and the stage where the citizens actually live in and use them after it is finished. In other words, design only becomes important in the first half of the process. Of course there is no doubt that Alexander emphasized the importance of the citizens nurturing their homes and towns while living in it. Still, the discontinuity that existed in the process was considerable.

In Pattern Language 1.0, designing was a one-time activity in the process of creating a town where things could rarely be undone after it was made. In contrast to this, Pattern Language 2.0 involved a system of updating the design. Software codes could be fixed and rewritten to be re-released as a new version. For this reason, pattern languages in the 2.0 stage can be said that it is continuously being designed.

The new 3.0 stage shows a new aspect. Pattern languages in this stage can be said that it is constantly being designed. Unlike architecture that has a concrete border that marks before and after the designing process, or software design where each version of the codes can be marked with the release of an update, human actions are something that is put into practice both constantly and continuously. Its details change nonstop from time to time, making it difficult to mark the beginning or the end of a certain phase. This phenomenon can be explained by the observation that the difference between designing and practicing a human action is hardly distinguishable. We practice to design and design to practice. The two actions coexist in our actions, together creating an endless and consistent process of designing.

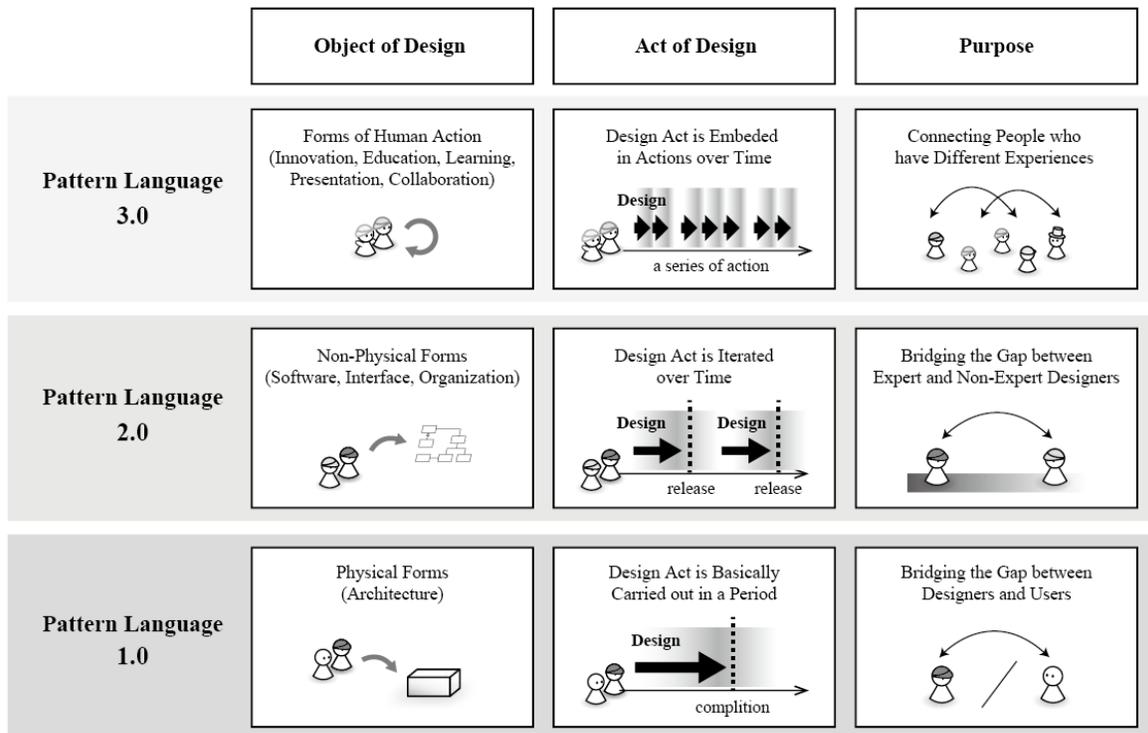

*Figure 3: Comparison among Generations of Pattern Languages*

*A New Purpose*

Alexander was critical on having a designer from outside to come in and design a community. His intent in writing out the design knowledge for architecture in a pattern language was so that the families could get involved in the designing process of their homes and towns. He believed that the pattern languages would serve as a lingua franca between the designers and users. His idea to bridge the two sides was the fundamental idea for the use of pattern languages in the Pattern Language 1.0 stage.

When the era shifted to Pattern Language 2.0, changes were seen to how the patterns were used. Its primary purpose became to be filling in the technical gap between the experts and the less experienced. Software designers who wished to improve their skills read the patterns to learn from the design knowledge of the more experienced programmers. In such a situation, "users" no longer exist. Both sides are now designers, and the patterns exist merely to lessen the distance between them.

The patterns take another turn in the 3.0 stage. They now are used to connect all kinds of people (actor) with all kinds of different experiences. 3.0 patterns help bring light to the less noticeable parts of a person's experience, so the person can reconsider the experience to talk and share about it to others. Stated simply, Pattern Language 3.0 becomes a medium of narrative and conversations between people.

Surprisingly, pattern languages function as media of narrative and conversations independent of the amount of experience or skill a person has. In addition, a person not need have any understanding about pattern languages nor background knowledge about the design knowledge written in it to use the pattern to their full advantage. Remember that pattern languages are media of narrative and conversations, and not tools to transfer high skills of profession. The ultimate goal for Pattern Language 3.0 is for people to be able to talk with each other about the different experiences they have based on their design knowledge, without having any difficulties.

## PATTERN LANGUAGES FOR BASIC SKILLS IN THE CREATIVE SOCIETY

As basic skills in the creative society, I focus on three creative acts: learning, collaboration, and presentation. For each of acts, I launched projects for creating pattern languages for them. The result of our projects is as follows (Figure 4): Learning Patterns, Collaboration Patterns, and Presentation Patterns.

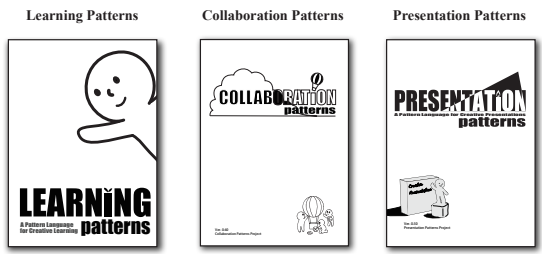

Figure 4: Three Pattern Languages for the Creative Society

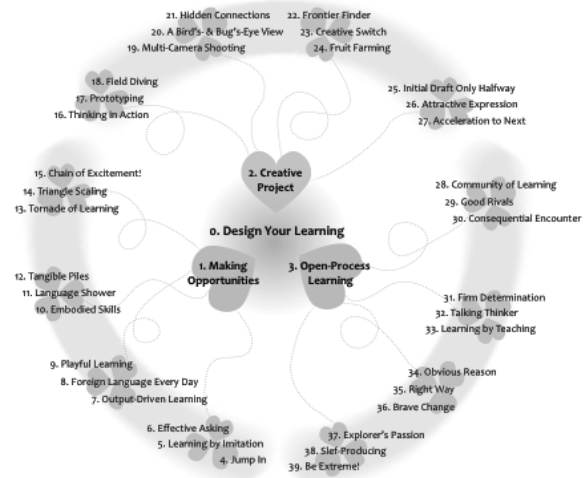

Figure 5: The Overview of the Learning Patterns

### Learning Patterns

Learning Patterns is a pattern language for creative learning, which consists of 40 patterns describing practical knowledge for problem finding and problem solving in learning are presented (Iba *et.al.* 2009; Iba and Miyake 2010; Iba and Sakamoto 2011). It provides an opportunity for learners to know ways that they have not experienced, but that are known as good ways, and also it encourages learners to talk about their ways for learning in their group or community.

The Learning patterns are organized in three layers according to the abstraction level (Figure 5). In the top layer, there is a root pattern: Design Your Learning. This pattern provides an introductory explanation about why and how to use the Learning Patterns. In the second layer, there are three fundamental patterns: Making Opportunities, Creative Project and Open-Process Learning. These patterns show essential minds for creative learning, summarizing more specific patterns in next third layer. In the third layer, there are thirty-six patterns as concrete `knack' of learning.

0. **Design Your Learning**
   Learn the way of learning from the Learning Patterns, which help you achieve good methods for learning.
1. **Making Opportunities**
   Make opportunities for learning by yourself, based on your interests.
2. **Creative Project**
   Launch your project and carry it out to improve your knowledge and skills.
3. **Open-Process Learning**
   Share your learning process and collaborate with others to deepen your and others' learning.
4. **Jump In**
   Jump into the new environment for your learning.
5. **Learning by Imitation**
   Begin with imitating the ways of the master to learn.
6. **Effective Asking**
   Clarify where you got stuck and then seek advice.
7. **Output-Driven Learning**
   Create an output in order to acquire knowledge and improve your skills.
8. **Foreign Language Every Day**
   Involve yourself in the situation of reading, writing and speaking in a foreign language in your daily life.
9. **Playful Learning**
   Take "play" into the process of learning.
10. **Embodied Skills**
    Continue practicing a skill again and again until you can use it unconsciously.
11. **Language Shower**
    Set up your environment where you always listen and read in the target language.
12. **Tangible Piles**
    Record the activities of your learning to improve your learning and reflect on your paths.
13. **Tornado of Learning**
    Collect information related to your interests like the vacuum of tornado.
14. **Triangle Scaling**
    Acquire knowledge related to what you want to understand, and you will understand it better.
15. **Chain of Excitement!**
    The strong emotion that you experience once you are able to grasp something and feel of accomplishment will motivate your learning.
16. **Thinking in Action**
    Deepen your thought process by making prototypes and doing fieldwork.

17. **Prototyping**
    Make some prototypes and consider how to make it better.
18. **Field Diving**
    Dive into the field and work with the people concerned while maintaining the viewpoint of an outsider.
19. **Multi-Camera Shooting**
    Collect a lot of information about targeted knowledge you wish to acquire, and understand it from various angles.
20. **A Bird's- & Bug's-Eye View**
    Take turns viewing the whole and then the detailed points.
21. **Hidden Connections**
    Explore hidden connections among things to attain inspiration.
22. **Frontier Finder**
    Grasp the frontier of the field, and then acquire the knowledge needed to reach that line.
23. **Creative Switch**
    Switch between two modes of logical and intuitive thinking.
24. **Fruit Farming**
    Do your best to put your idea into shape as a seed, and then nurture it.
25. **Initial Draft Only Halfway**
    After finishing an initial draft, brush it up with an objective view, while considering whether or not readers can easily understand.
26. **Attractive Expression**
    Find better ways of expression for attracting others.
27. **Acceleration to Next**
    Set the next goal and pass through the current goal without slowing down.
28. **Community of Learning**
    Build a community of learning with people who share similar interests.
29. **Good Rivals**
    Make good rivals and inspire each other.
30. **Consequential Encounter**
    Find people who have similar interests as yours by getting involved in the field you are interested in.
31. **Firm Determination**
    Firmly determine what you are going to do, and set up the environment for concentrating on it.
32. **Talking Thinker**
    Explain what you think to improve your idea.
33. **Learning by Teaching**
    Teach others your knowledge while considering their levels, and you can gain an understanding on various levels.
34. **Obvious Reason**
    Confirm the meanings of your assumptions by questioning yourself again.
35. **Right Way**
    Consider whether your current way is actually correct or not; then quickly change your approach as necessary.
36. **Brave Change**
    Throw away previous themes or approaches to achieve a wider view for the future.
37. **Explorer's Passion**
    Choose a topic that you can be passionate about– something for which you feel "love" or "rage."
38. **Self-Producing**
    Design a concrete plan to achieve your goal while inventing your future.
39. **Be Extreme!**
    Think strategically where you can/want to be distinguished and just do it in order to greatly differentiate yourself from others.

**Collaboration Patterns**

Collaboration Patterns is a pattern language for creative collaboration, which consists of 34 patterns describing practical knowledge for performing collaborations (Iba and Isaku 2013). A creative collaboration here is teamwork that creates new values that would change the world by producing an emergent vigor that cannot be reduced to any one member of the team, with those who you can grow together by enhancing one another.

The Collaboration Patterns consist of 34 patterns (Figure 6). At the center of the pattern language sits pattern No.0 Creative Collaboration, and the three ultimate goals for such a collaboration: Mission for the Future (No.1), Innovation of the Ways (No.2), and Create a Legend (No.3) surround this. Subsequent patterns are grouped into three main categories. Each of the three categories are started off by the three main elements of a Creative Collaboration: Spiral of Growth (No.4), Vigor of Emergence (No.14), and the Power to Change the World (No.24). The first group, patterns No.4 - No.13, deals with how the team should operate. Patterns No.14 -No.23, describes tips for creation. The final group of patterns, patterns No.24 - No.33, gives advice on how the project can achieve even further when the project is near its finish. These patterns, through its interactions, help teams achieve a creative collaboration.

0. **Creative Collaboration**
    Create new values that would change the world by producing an emergent vigor that cannot be reduced to any one member of the team, with those who you can grow together by enhancing one another.

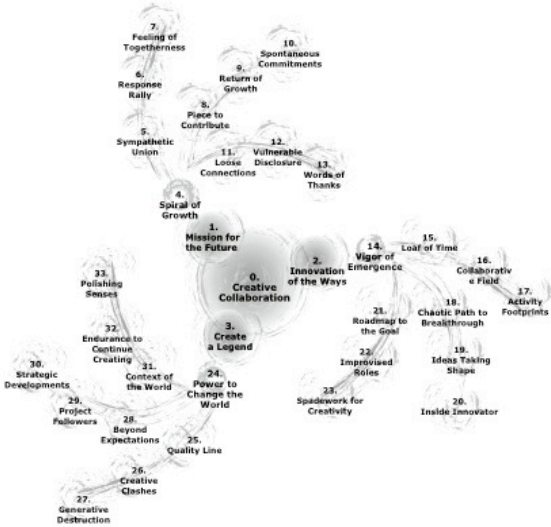

*Figure 6: The Overview of the Collaboration Patterns*

1.  **Mission for the Future**
    Have an image of how the future must be, and start working with the sense of mission that you must be the one to make this future a reality.
2.  **Innovation of the Ways**
    Pay attention to the creation process of the project. Invent new ways of creating and put it to practice.
3.  **Create a Legend**
    Set towards a project that will become a part of history with the mindset of changing the world.
4.  **Spiral of Growth**
    Get stimulated from the efforts of other team members, and then stimulate them. Work on the project as a team and grow together dynamically.
5.  **Sympathetic Union**
    Search for members who you can share a common vision towards the future.
6.  **Response Rally**
    Always give a response to what members say, no matter how small the response is.
7.  **Feeling of Togetherness**
    Beside the individual tasks, organize opportunities for the team to share common experiences working on the project.
8.  **Piece to Contribute**
    Think for yourself how your specific knowledge's and skills can be used for the project.
9.  **Return of Growth**
    Identify the skills and experiences you want to gain through the project, and clarify the meaning for you to be a part of this project.
10. **Spontaneous Commitments**
    Have the feeling that the project belongs to you, and actively search for tasks that could be done for the project.
11. **Loose Connections**
    Arrange a system where members can loosely communicate outside project hours and have a grasp of how each member is doing when project meetings start.
12. **Vulnerability Disclosure**
    Relating to the project or not, disclose your troubles and worries that you have to your team members.
13. **Words of Thanks**
    Reconsider the environment that you are in, and show thankfulness to the support that you are receiving.
14. **Vigor of Emergence**
    Create an environment that encourages members to give the slightest of ideas, and trigger emergent chains of ideas in the team.
15. **Loaf of Time**
    Schedule plenty of continuous time available to work on the project.
16. **Collaborative Field**
    Create with the team a place where members can have fun working creatively; free to concentrate to the full extent.
17. **Activity Footprints**
    Share the progress of each person with the whole team and make it possible to track down past versions of everyone's work in chronological order.
18. **Chaotic Path to Breakthrough**
    Recapture the current situation as a chance to innovate new ways; stay where you are and think through the situation thoroughly.
19. **Ideas Taking Shape**
    Visually shape your idea so others can see while you explain.
20. **Inside Innovator**
    When you have a good idea, nurture its growth by gradually finding people who understand the idea and by getting them involved.
21. **Roadmap to the Goal**
    Talk together as a team on how the project will run, and set milestones for the project by calculating backwards from the deadline.
22. **Improvised Roles**
    Take into account the progress of the project, look at the situation of other members, and do what must be done in top priority.
23. **Spadework for Creativity**
    Prepare small surprises that would liven up members and trigger new ideas in them.
24. **Power to Change The World**
    Keep a questioning mind and keep asking if your project has the "Power to Change the World."

**25. Quality Line**
Set high goals for the level of quality to reach, and check how much the current level differs from it over and over again.
**26. Creative Clashes**
Tell all of your opinions and have earnest discussions to reach true quality.
**27. Generative Destruction**
Take apart what the team has now and re-create the product from the start.
**28. Beyond Expectations**
Think what kinds of expectations the recipient will have, and make a product that goes beyond those expectations.
**29. Project Followers**
Let the project have an original view of the world that would excite people and make them want to become fans of it.
**30. Strategic Developments**
Think of multiple ways the project could be used besides its original purpose.
**31. Context of the World**
Redefine the significance and the value for the team to work on this project by taking into account the context of the world.
**32. Endurance to Continue Creating**
Train yourself regularly for a strong body and mind that could bear the hardships of creation.
**33. Polishing Senses**
Obtain a better sense of quality by enjoying sensible work by others.

**Presentation Patterns**

Presentation Patterns is a pattern language for creative presentations, which consists of 34 patterns describing practical knowledge for designing presentations (Iba *et .al.* 2012). A creative presentation is an imagination provoking presentation with the presenter's feelings condensed into it that invites the audience to new findings. Although we use the word "presentation," it can be applied to all kinds of activities of representation, including public speaking, performance of music, drama, and dance.

The Presentation Patterns consists of 34 patterns (Figure 7). In the center of the patterns sits pattern A Creative Presentation (No.0). Three of the main patterns: Main Message (No.1), A Touching Gift (No.2), and Image of Success (No.3) surround this. Subsequent patterns are grouped into three categories. The first group, patterns No. 4 – No.12, deals with the contents and the expressions of the presentation. Patterns No.12 –No.21 consider how to make your presentation appealing to the audience. How you should act once you are on stage is discussed in the final set of patterns, No.22 – No.30. The Presentation Patterns then wraps up with the final three patterns:

Pursuit of Uniqueness (No.31), Aesthetics of Presenting (No.32), and Be Authentic! (No.33). By combining these patterns according to your needs, you can give a truly creative presentation.

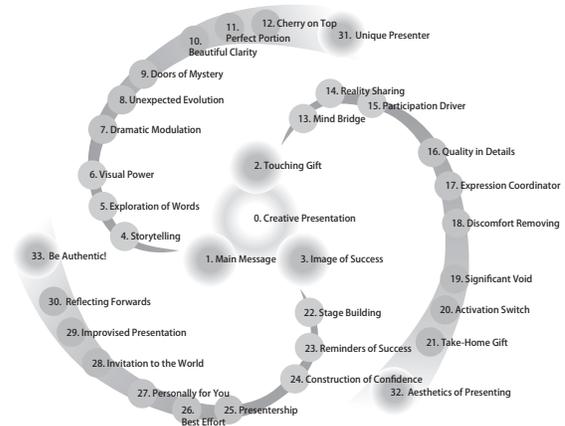

*Figure 7: The Overview of the Presentation Patterns*

**0. Creative Presentation**
Treat your presentation not as just a chance to explain your idea, but as a chance for creation. Work towards the audience to trigger new findings in them.
**1. Main Message**
Extract the one most important message, and create your presentation around that idea.
**2. Touching Gift**
Focus on who the audience is and think how to specifically make them impressed with your message.
**3. Image of Success**
Have an image of success for your presentation.
**4. Storytelling**
Create an attractive story with the information you have.
**5. Exploration of Words**
Search for words and expressions that both the presenter and the audience feel attractive.
**6. Visual Power**
Take advantage of a visual figure that expresses the information.
**7. Dramatic Modulation**
Make a modulation in your presentation by creating a difference in your tone when telling the Main Message (No.75) against the other parts.
**8. Unexpected Evolution**
Intentionally swerve the Storytelling (No.78) from the audience's expectations to add an interesting and unpredictable turn.
**9. Doors of Mystery**

Design the structure of the presentation so that it appeals to the curiosity of the audience and continuously drives their interest.

10. **Beautiful Clarity**
    Brush up your presentations so it has a good balance between clarity and allure.
11. **Perfect Portion**
    Make sure your presentation has just the right amount of information at various levels of the presentation.
12. **Cherry on Top**
    Make extra improvements in not the contents of the presentation, but in its expressions.
13. **Mind Bridge**
    Use metaphors and specific examples to help explain to the audience.
14. **Reality Sharing**
    Make an opportunity within the presentation for the audience to actually experience first-hand the sensation you want to share.
15. **Participation Driver**
    Make an opportunity for the audience to participate in your presentation.
16. **Quality in Details**
    Taking into account the presentation as a whole, work on fixing the small details.
17. **Expression Coordinator**
    Watch other people present, and take in any techniques that you think would help your presentation skills.
18. **Discomfort Removing**
    Make an opportunity for yourself to notice your habits.
19. **Significant Void**
    Purposely leave out some information so the listeners can use their imagination to fill in the missing parts.
20. **Activation Switch**
    Include in your presentation a chance for the listeners to think about their own opinions, and then show them the path for the next step.
21. **Take-Home Gift**
    Hand out a gift for the audience to take home that would remind them of the details of the presentation.
22. **Stage Building**
    Treat facilities and equipment as part of your presentation, and check thoroughly for any problems and adjust them beforehand.
23. **Reminders of Success**
    Frequently remind yourself with your Image of Success (No.77), whether verbally or visually, to refresh your mind as needed.
24. **Construction of Confidence**
    Remind yourself of all the time and effort you have put into this presentation, stack them up and build your self-confidence.
25. **Presentership**
    Act as if you are part of the presentation.
26. **Best Effort**
    Make no excuses and give the best presentation you have at the point.
27. **Personally for You**
    Intentionally make eye contact with the audience in effort to give the speech to each and every one of them.
28. **Invitation to the World**
    Have an alluring "world" of your presentation, let the audience have a glimpse of that "world", and then guide them into it.
29. **Improvised Presentation**
    Have a repertoire of speeches that you can put together and improvise based on the reactions of the audience.
30. **Reflecting Forwards**
    Evaluate your performance on the presentation that you gave through self-reflection and the reaction of others.
31. **Unique Presenter**
    Be aware of the differences you have from other presenters, and pursuit your originality.
32. **Aesthetics of Presenting**
    Continue your pursuit for beauty in your presentations, and build your own values on aesthetics.
33. **Be Authentic!**
    Live your way of Life.

## SYSTEMS ANALYSIS OF THE FUNCTUIONS OF PATTERN LANGUAGES

### Analyzing based on Social Systems Theory

Social systems theory takes communication as a whole of three choices, "information," "utterance" and "understanding" (Luhmann 1984). Communication as a whole is raised as one "meaning," thus it cannot be resolved. These "selection of information," "selection of utterance" and "understanding information and utterance" occurs at the same time, therefore it can be said that communication is the phenomenon of emergence. This way of understanding communication is not from a perspective of human being but from a perspective of communication itself. Thus, the first step is to understand that "communication is to select," and then there would be rooms for creativity in communication. It is because creativity can be incorporated into the contingency of selectivity. Additionally, selectivity in this sense does not mean to select from given choices, rather it means to create choices by itself at the same time. As Luhmann states that "Communication grasps *something* out of the actual referential horizon that it itself constitutes and leaves *other things* a side." (Luhmann 1984).

In order to understand the mechanism of pattern language and the way it functions, this study borrows the idea of the social systems theory. First, as it says, pattern language functions as "language." Language is "the medium that increases the understandability of communication beyond the sphere of perception" (Luhmann 1984), it enables to understand what others are thinking in the communication of the social system. It also has certain role in structuring consciousness of the psychic system. Language has been shrinking its own multiplicity due to the limits of possible combinations between symbols, "this concerns a very special technique with the function of *extending* the repertoire of understandable communication *almost indefinitely in practice* and thereby guaranteeing that almost any random event can appear and be processed as *information*" (Luhmann 1984). Language couples the social system and psychic system while it plays its roles. Thus, the social system and psychic system do not merge in the process, however, they can affect on each other indirectly through pattern language.

Addressed pattern language affects on the psychic system of others, and helps structuring it, thus even beginners can experience the nexus of consciousness as experts do. It also plays a role to reduce the composition of the world by making a part of possible options of a solution conspicuous. By making a pathway on the nexus of consciousness, each individual becomes able to think efficiently. Therefore, by helping to structurize the nexus of consciousness, pattern language is capable of encouraging creative thinking and creative action.

**Analyzing based on Creative Systems Theory**

In the Creative Systems Theory (Iba 2009), the creative process consists of a sequence of discoveries, which include problem finding, problem solving, observation, hypothesis formation, method selection, practice, and interpretation. The creative process does not follow deterministic laws, but it also does not happen at random. Rather, it includes contingency. The creative process is, so to speak, autonomous and therefore historical. In order to formulate this kind of processes, we would like to apply autopoietic systems theory. The creative systems theory describes how creation is possible. This attempt is done without psychological reduction, as most creative researches do, nor sociological reduction, as most studies of collaboration do.

In order to describe the creative process as a sequence of discoveries, we would like to suggest that creativity is an autopoietic system whose element is discovery. In creative systems, discovery is produced by discovery based on on-going creation.

The discovery is a momentary element that has no duration, so it must be constantly reproduced in order to realize the creative system. Element, discovery, is an emergent unity constituted in the system, therefore the system cannot receive discoveries from its environment or output discoveries to its environment. In this sense, the kinds of discoveries that are made depend on the ongoing system. Thus the creative system is operationally closed.

From the viewpoint of element constitution, discovery emerges from the synthesis of the three-part selection: *idea*, *association*, and *finding* (Figure 8). It is required for the emergence of discovery that all of these selections are occurred. What we should emphasize here is that idea exists only inside the system. It other words, idea is meaningful only for ongoing creation. Outside the creation, one can no longer call it "idea." In this sense, idea cannot exist "out there" alone. In the same way, association can exist meaningfully only inside the system. It is just association to ongoing creation. Consequence occurs only as the combination of idea and selection, therefore it also can exist only inside the system.

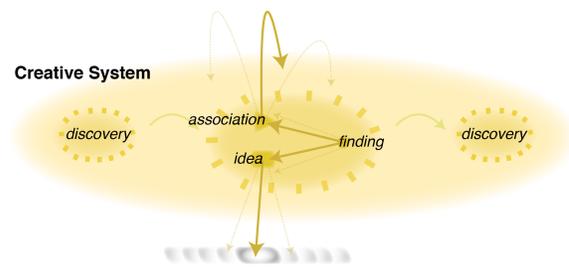

*Figure 8: Element Constitution in Creative Systems*

Pattern Languages urge the chain of discovery to happen in people. There are multiple aspects of Pattern Languages that make this chain happen.

First, the *idea* of the pattern itself contains the potential for giving its users new discoveries. When users first read the pattern, the realization of the new perspective will allow the users to view situations from a different angle than usual. For example, the pattern *Chaotic Path to Breakthrough* from the *Collaboration Patterns* tells readers to value the stagnations that occur in the creation process, and to think through it carefully by recapturing it as a chance for breakthrough that is about to come. This pattern gives people a new perspective to look at such silence and confusion that occurs in projects, and the person stuck in the chaos may try to think out through the situation instead of just giving up and searching for a different approach. Similarly, the *Presentation Patterns* also contain patterns that simply give people new perspectives for creating a presentation. *Exploration of Words*, *Doors of Mystery*,

*Triggering Blanks,* and *Take-Home Gift* are of this type.

The second way a pattern can help the chain of discoveries in a person is by giving people a new way with deal with a *problem*. People may be aware of a situation and know that it is a problem, but not know how to cope with it. These patterns would give people a path to escape such a *problem* through its *solution and actions*. *Response Rally* from the *Collaboration Patterns* is an example of this type. We often find ourselves in situations where we know we should respond to an email but don't know what to say, and end up not responding at all. This would result in delays in the project, and the motivation of the other person would drop if she did not get any responses form anyone. The pattern gives people suggests a solution on how to deal with the situation by saying to show a response no matter how small and short it may be. When read the person is met with a finding that may have not been noticed without the pattern.

The *Problem* of a pattern can also provide people with a new discovery. We may become so used to a problematic situation that we usually do not notice that it is a problem. Once the *problem* is noticed, then conscious efforts can be made to avoid the *problem* from happening. For example, the pattern *Return of Growth* from the *Collaboration Patterns* has the problem written, "Your motivation to contribute to the project will not last in the long run if thoughts are centered only on contributing to the project". When read, the problem would make the person think if there is anything else needed for them besides the usual contribution. *Best Effort* from the *Presentation Patterns* is another example of this type. Its *problem* states, "You tend to make excuses for the parts you don't have as much confidence in, only to make the presentation even worse". This too is usually unnoticeable without the pattern, and gives the person a chance to reflect on their presentation to see if they have the problem.

As shown, each Pattern gives people a new discovery through its *idea, solution and actions*, or its *problem*. These patterns are both explicitly and implicitly related to other patterns in the language. When a person is met with a discovery with one pattern, he can also trace these relationships to see if any of the other patterns would also bring them new findings. Since these patterns are written in small units, and many patterns exist within a pattern language, they become a good tool for bringing a chain of discovery to people.

## DIALOG WORKSHOP WITH A PATTERN LANGUAGE

To introduce these pattern languages to a certain community, I usually held workshops using the pattern languages, where participants talk about their experiences with each other using the pattern language. The enjoyment in the workshop will help participants to think, learn, and talk about the patterns. The participants learn and think to use the pattern language with enjoyment.

The workshop first asks participants to read the all patterns beforehand, and list the patterns that they have experienced already in their learning career. They would also make a list of five patterns in which they wish to use in their future learning. With the list of the five patterns in their hands, the participants would walk around at the workshop searching for other participants who have already experienced one of the patterns on the list. Once they find a match, the one who has experienced the pattern would talk about the experience, and the other person would take notes (Figure 9).

Although a simple structure, the workshop fosters unexpected excitement among participants. In daily conversations, we tend to think the episodes of our learning experiences are pointless to talk about. These seemingly useless episodes are given meaning by applying it to one of the patterns, lowering the bar to share it. Plus, the positive feedback from the other participants who wish to hear the story would encourage them to talk to more people about their experiences. The positive chain of episodes and feedback adds up to warm up the atmosphere and bring excitement to the workshop.

By using pattern languages as media of narrative and conversations we can become conscious of the parts of our experiences we normally do not notice, and inspect it through conversations. Stories of the experience will not only stimulate conversations, but also would create an opportunity to reflect on one's own learning styles. Fun and live conversations about each other's experiences would arise and foster their learning. Consequently, the 3.0 patterns would cause the "design knowledge" to be learned intrinsically. This is a remarkable difference from the Pattern Language 2.0 stage where the users of the patterns had to read and try to understand the patterns by themselves.

We've already more than 30 workshops for various targets, in various domains, and in various places with using the Learning Patterns, Presentation Patterns, or Collaboration Patterns; The total number of participants are more than 3,500 people (Figure 10, 11).

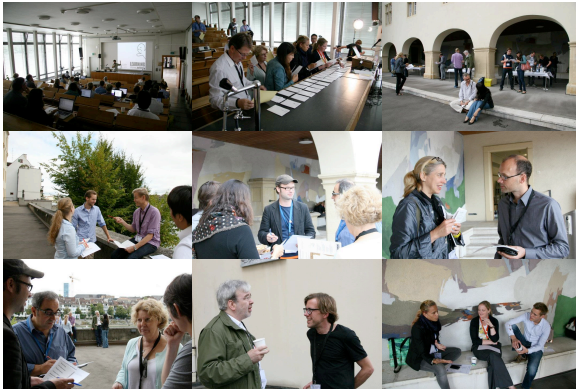

*Figure 9: Dialog Workshop with a Pattern Language (Iba 2011b)*

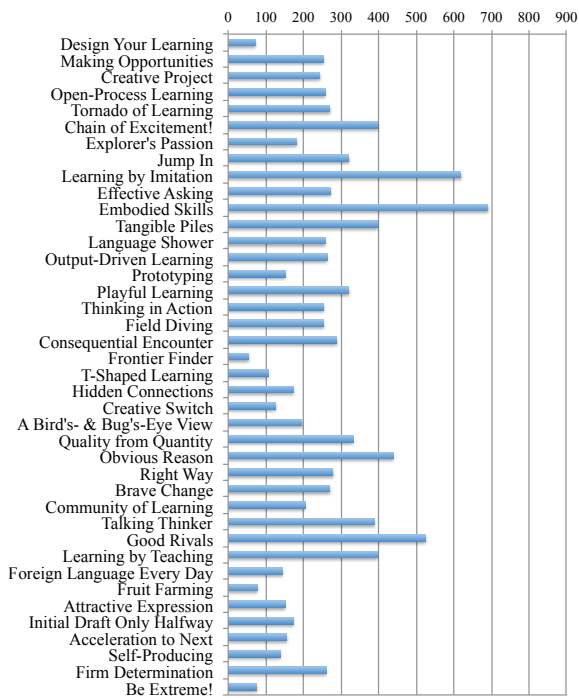

*Figure 10: The distribution of the experience of all participants for all patterns.*

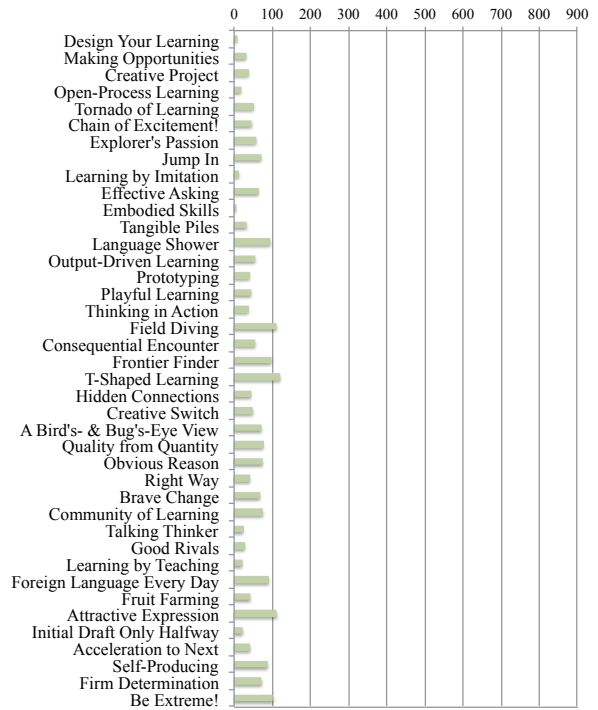

*Figure 11: The distribution of the patterns that participants want to gain.*

## CLOSING - CREATING THE CREATORS

I have talked about the role of Pattern Languages as a communication media or a Creative Media. In addition to these aspects, Pattern Languages support humans by becoming what can be called Glasses of Recognition. These glasses give its users an additive perspective to the world around them. With the patterns in their minds, they can cut out otherwise unnoticed events from their environments. The chain of these small findings are what helps people become ore creative.

The Collaboration Patterns for example becomes a pair of Glasses of Recognition to look at teams. By looking at their own teams through the lenses of the patterns, team members can notice what their team is already able to do that is making their team successful, or what their team can additionally do to make the team better, in terms of the Collaboration Patterns.

Off from the glasses metaphor, the patterns eventually assimilates and becomes a part of the body. After a while, the users will be able to notice the patterns without consciously putting on the glasses. In this sense, it can be said that Pattern Languages are media which will eventually vanish.

The important part of these Glasses of Recognition, is that they are not to be just received from others and consumed, but they can be created by anyone for their own use. In the following years, while there are still severe needs for the creation of new Patterns Languages in various fields, even more efforts must be put into the education and support of pattern creators who then would be able to make Patterns by themselves. This movement will stimulate the creative actions of the people living in the creative society, and these people will become ready for the wave of the new movement about to come in the next few decades.

## ACKNOWLEDGEMENT

I want to thank to Taichi Isaku for helping writing this paper, and all members of the Learning Patterns Project, Collaboration Patterns Project, and Presentation Patterns Project.